\documentclass[onecolumn,preprintnumbers,superscriptaddress,nofootinbib,aps,prd,floatfix]{revtex4}

\usepackage{footmisc,multirow}
\usepackage{subfigure,color}
\usepackage{amsmath,amsfonts,amssymb,slashed,enumerate,mathrsfs}

\usepackage{graphicx}

\newcommand{\be}{\begin{eqnarray*}}
\newcommand{\ee}{\end{eqnarray*}}

\newcommand{\bee}{\begin{eqnarray}}
\newcommand{\eee}{\end{eqnarray}}
\newcommand{\beeq}{\begin{equation}}
\newcommand{\eeeq}{\end{equation}}

\newcommand{\tev}{{\text{TeV}}}

\preprint{IPPP/15/47} \preprint{DCPT/15/94}

\begin{document}

\title{Augmenting the diboson excess for Run 2}

\begin{abstract}
The ATLAS collaboration recently reported an excess of events in the high invariant mass tail
of reconstructed di-boson events. We investigate their analysis and point to possible subtleties and improvements 
in the jet substructure implementation and data-driven background estimates. 
\end{abstract}

\author{Dorival Gon\c{c}alves} \email{dorival.goncalves@durham.ac.uk}
\affiliation{Institute for Particle Physics Phenomenology, Department
  of Physics,\\Durham University, Durham, DH1 3LE, United Kingdom\\[0.1cm]}

\author{Frank Krauss} \email{frank.krauss@durham.ac.uk}
\affiliation{Institute for Particle Physics Phenomenology, Department
  of Physics,\\Durham University, Durham, DH1 3LE, United Kingdom\\[0.1cm]}

\author{Michael Spannowsky} \email{michael.spannowsky@durham.ac.uk}
\affiliation{Institute for Particle Physics Phenomenology, Department
  of Physics,\\Durham University, Durham, DH1 3LE, United Kingdom\\[0.1cm]}

\maketitle

\section{Introduction}
\label{sec:intro}

Recently, both ATLAS and CMS reported excesses in the high invariant-mass tail of di-jet 
events when subjet-based reconstruction techniques were applied, suitable for example 
for highly boosted gauge bosons~\cite{Aad:2015owa,cms}.
While the CMS observation~\cite{cms} of an excess of around $2\sigma$ with respect to the
Standard Model expectation about two years ago went relatively unnoticed by the wider
high-energy physics community, the ATLAS publication~\cite{Aad:2015owa} reporting an 
excess of about $2.5\sigma$ over background estimates triggered a flurry of mainly 
theoretical papers aiming at an explanation of this excess.  Clearly, such an excess 
in the mass range of about 2 TeV, as reported, may easily be the first glimpse of new 
physics beyond the Standard Model and it is therefore not surprising that practically
all publications to date interpret the excess as the intriguing discovery of a TeV-scale 
resonance decaying predominantly to the gauge bosons of the weak interaction.  This
of course is a scenario predicted in many extensions of the Standard Model of
particle physics~\cite{bsm}.  

However, many of the searches for these high-mass objects heavily rely on good theoretical 
and experimental control of often novel reconstruction techniques in highly exclusive
and hitherto unprobed regions of phase space.  In addition, as the exciting signals for 
new physics often manifest themselves through a very small number of events only, a 
precise understanding and book-keeping of all possible backgrounds is of paramount
importance to the full appreciation of the statistical significance of any excess or
lack thereof.  Hence, before the incoming 13 TeV data, it is important to understand the techniques employed by such searches in detail and provide constructive assessments of the potential for biases or improvements in the analysis strategy.

As the latest analysis by ATLAS on this subject~\cite{Aad:2015owa} yields the best 
sensitivity and largest statistical significance, and because it is relatively
well documented, we will focus on it throughout this short paper.  This search is based 
on the assumed decay of a heavy resonance in two weak gauge bosons and, consequently, 
on looking for excesses in the invariant mass distribution of two hadronically decaying 
gauge bosons, thus using the larger hadronic branching ratio.  This was realised by 
analysing events based on a fat-jet selection with the Cambridge/Aachen (C/A) 
algorithm~\cite{Wobisch:1998wt}, demanding two $R=1.2$ jets with a minimal transverse 
momentum of $p_T^{\rm min}=540$ GeV.  The reconstruction of the gauge bosons relied
on a combination of a jet-mass cut around the masses of the weak gauge bosons
and grooming techniques, where a modified version of the BDRS reconstruction 
technique~\cite{Butterworth:2008iy} was employed, a method initially designed for the 
reconstruction of a Higgs boson with $p_{T,H} \geq 200$ GeV.  Eventually, to improve 
on the separation of signal and background, cuts were applied on the momentum ratios 
of subjets, the number of charged particles within a subjet and the mass of the 
reconstructed gauge bosons.  After recombining the four-momenta of the two reconstructed 
gauge bosons an excess was observed in the mass range $1.9 \leq m_{VV} \leq 2.1$ TeV over 
the data-driven (fitted) background estimate, mainly driven by the QCD background.

Following and re-implementing the analysis steps as described in~\cite{Aad:2015owa}, 
we find several points that could potentially reduce the statistical significance of 
the observed excess.  While our findings may possibly be attributed to our lacking 
understanding of subtle and possibly internal details of the analysis, we feel
that they warrant a more detailed discussion.  Broadly speaking, our findings fall
into two categories: one, discussed in Sec.~\ref{sec:Substructure}, is related to the 
reconstruction method used for the gauge bosons, which open a number of potential 
pitfalls and some room for future possible improvements.  The other class of comments
relates to the modelling of backgrounds by data-driven methods in general, and in 
Sec.~\ref{sec:datadriven} we will point to backgrounds that contribute predominantly 
in the tail of the reconstructed $m_{JJ}$ distribution, thereby evading the estimate on 
which the analysis here is based.  We finish the paper with a short summary of possible 
lessons for Run II.  

\section{Jet Substructure}
\label{sec:Substructure}

The study of jet substructure techniques has received a lot of attention over the last years, 
with the high-energy community increasingly appreciating the benefits of producing electroweak-scale 
resonances beyond threshold \cite{Altheimer:2013yza}. Particularly, when a TeV-scale resonance $Y$ decays into 
electroweak-scale resonances $X$ with a large branching ratio into quarks, the decay products 
of $X$ will be confined in a small area of the detector, i.e. a small jet, and jet-substructure 
techniques will become unavoidable.

Dedicated reconstruction techniques for the Higgs boson, top quarks,  and electroweak gauge 
bosons \cite{Butterworth:2008iy,tagger,Schaetzel:2013vka,Spannowsky:2015eba} have been designed and tested by both multi-purpose experiments. However, 
despite ongoing efforts an analytical understanding of these tools has only progressed for simple
 jet substructure observables \cite{Dasgupta:2013ihk, calc} and has yet to be achieved for most of the high-performance taggers.

Hence, apart from theoretical insights, for an adequate application of these tools a detailed
understanding of their limitations is crucial, as the flexibility, purpose of and required input to 
each of the taggers may differ. In the reconstruction of highly boosted resonances the resolution
of the input objects used is of crucial importance. 

As input to the jet algorithms ATLAS is relying on topoclusters, a combination of cells 
from the hadronic and the electromagnetic calorimeter. The cell size of the ATLAS hadronic
 calorimeter is $0.1 \times 0.1$ in $(\eta, \phi)$ and topological cell clusters are formed around 
 seed cells with an energy $|E_\mathrm{cell}| > 4 \sigma$ noise \cite{Aad:2012vm}. Two particle jets leave
  distinguishable clusters if each jet hits only a single cell and the jet axes are separated by at 
  least $\Delta R = 0.2$, so that there is one empty cell between the two seed cells. It is possible to obtain smaller topoclusters by focusing on the electromagnetic calorimeter and tracks, hereby trading energy resolution against an improved spacial resolution for the jet constituents \cite{Katz:2010mr,Schaetzel:2013vka}. The latter path was chosen in this analysis (see topoclusters in Fig.~\ref{fig:event}).

\subsection{Kinematics of a possible signal}

\begin{figure*}[ht!]
\begin{center}
\includegraphics[width=0.5\columnwidth]{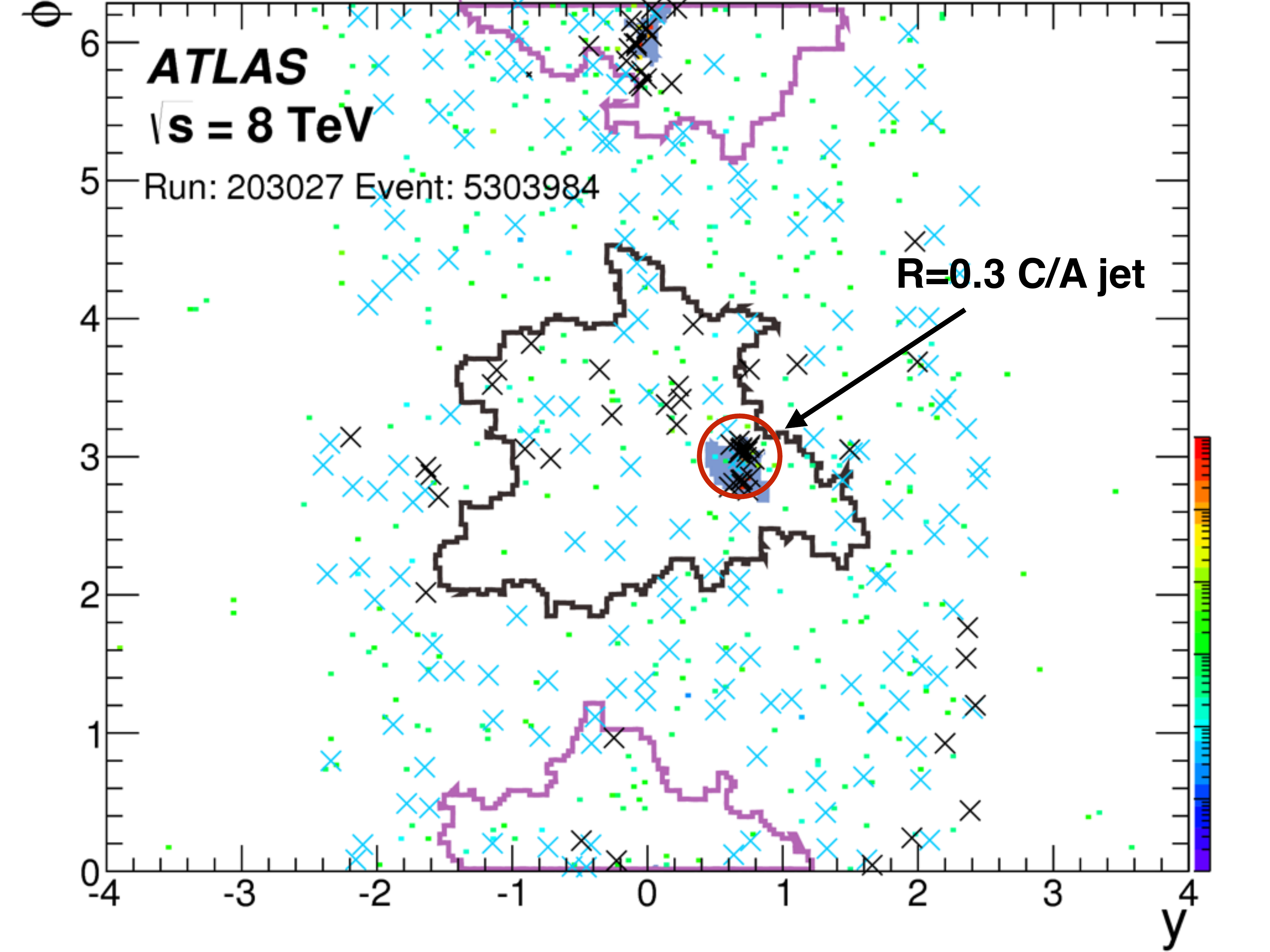}
    \parbox{.8\textwidth}{\caption{Event display taken from \cite{extMat}. Colored boxed correspond to topoclusters, 
blue crosses represent tracks not associated with the primary vertex, black crosses represent
tracks associated with the primary event vertex and the reclustered jets kept after the filtering
algorithm. For this event one finds $p_{T,J_1} = 999.3$ GeV, $p_{T,J_2}=999.3$ GeV and $m_{JJ} = 2068.6$ GeV.
 \label{fig:event}
}}
\end{center}
\end{figure*}

The analysis of consideration \cite{Aad:2015owa} is designed as a ``bump hunt" for a heavy
TeV-scale resonance decaying into W and/or Z bosons. While ATLAS is apriori not searching
for a resonance with a specific mass, the fat jet trigger cuts of $p_{T,J} \geq 540$ GeV and the
kinematic endpoint of the $m_{JJ}$ distribution limit the resonance mass range that can be
probed to $1.0 \lesssim m_Y \lesssim 2.1$ TeV. The opening angle between the quarks 
produced in the decay of W and Z bosons with large transverse momentum can be estimated by
\begin{equation}
\Delta R _{q \bar{q}} \simeq \frac{2 m_V}{p_{T,V}},
\end{equation}
where $m_V$ is the mass of the gauge boson and $p_{T,V}$ its transverse momentum. Hence, 
over the whole mass range of $Y$, for central production of the gauge bosons, the energy released 
in its decay is captured in two small spots of the detector with {\it diameters} in the range of 
$0.16 \lesssim \Delta R _{q \bar{q}} \lesssim 0.33 $, see also Fig.~\ref{fig:mjj} (left). This 
means, for most of the relevant mass range of $Y$ all energy of the W or Z decay products is
 contained in a very small area of the detector, i.e. a small number of topoclusters. Following this simple argumentation a fat jet cone size of $R=1.2$, as applied in \cite{Aad:2015owa}, is not motivated with a possible TeV-scale resonance
 decay to gauge bosons in mind. We give a graphic example for this scenario using an ATLAS 
event display in Fig.~\ref{fig:event}: While both fat jets together cover almost the entire detector 
in the central part, the region of interest, where the radiation can be found to reconstruct a heavy 
resonance, can be covered by two jets with radius R=0.3. The choice of a large jet radius can
increase the probability of underlying event, initial state or pileup radiation to distort the
reconstruction of the boosted resonances. As a result, jet grooming procedures that remove 
uncorrelated soft radiation have to be applied.

\subsection{Gauge boson reconstruction}

After reconstructing  two C/A fat jets with $R=1.2$, in \cite{Aad:2015owa} a grooming 
procedure is applied to remove soft, uncorrelated radiation from the jet. In \cite{Dasgupta:2013ihk} it was shown that grooming procedures can shape the backgrounds of QCD jets for small values of $\rho = m_J^2 / (p_{T,J}^2 R^2)$, where $m_J, p_{T,J}$ and $R$ are the fat jets mass, transverse momentum and jet radius respectively, implying that in general the fat jet radius should be adjusted for the mass scale of interest. However, by ATLAS choosing the so-called mass-drop tagger \cite{Butterworth:2008iy} to groom the fat jet, a sculpturing of $d\sigma/d\rho$ can be avoided which could have led to an increased fake rate for high-$p_T$ jets and hence a bump in the $m_{VV}$ distribution for large invariant masses. Still, since $d\sigma/d\rho$ has only been calculated to modified-leading-log accuracy, changing the grooming procedure and adjusting $R$ such that $0.01 \lesssim \rho \lesssim 0.1$ always, might improve confidence in the robustness of the chosen grooming procedure further.

The grooming procedure applied reverses the sequential jet recombination algorithm. When examining the pairwise
combinations used to construct the jet in reversed order, at each step the lower-mass 
subjet is discarded and the higher-mass subjet is kept for further declustering. 
Declustering stops when a pair $\left \{ j_1,j_2 \right \} $ is found that satisfies
\begin{equation}
\label{eq:ycut}
\min(p_{T,j_1}, p_{T,j_2}) \frac{\Delta R_{(j_1,j_2)}} {m_{j_1+j_2}} \geq \sqrt{y_f},
\end{equation}
with $\sqrt{y_f} =0.2 $. 

In this analysis the so-called mass drop condition of \cite{Butterworth:2008iy} was not 
imposed. The combination of mass-drop condition and y-cut was proposed to indicate the stage of
the jet recombination where the two subjets containing each a bottom jet were merged.
For a Higgs boson produced in associated production with a gauge boson, even after 
requiring $p_{T,H} \geq 200$ GeV, the angular separation of $p_{T,j_1}$ and $p_{T,j_2}$
would still be fairly large, e.g. $\Delta R \simeq 1.0$. Hence, at this point one would 
find two fairly massive, large, irregularly shaped subjets. To improve the mass resolution 
for the reconstructed Higgs further the authors of \cite{Butterworth:2008iy} propose
to use the constituents of $j_1$ and $j_2$ and recluster them with C/A $R\simeq 0.2$,
this step was called filtering. Jets with at least 2 or 3 filtered subjets were kept and
recombined to the reconstructed Higgs mass, softer subjets were discarded. The 
purpose of each step is to reduce the active area of the jet while keeping the relevant 
 parts of the jet to achieve an optimal Higgs mass reconstruction.
\\

\begin{figure*}[t!]
\begin{center}
\includegraphics[width=0.41\columnwidth]{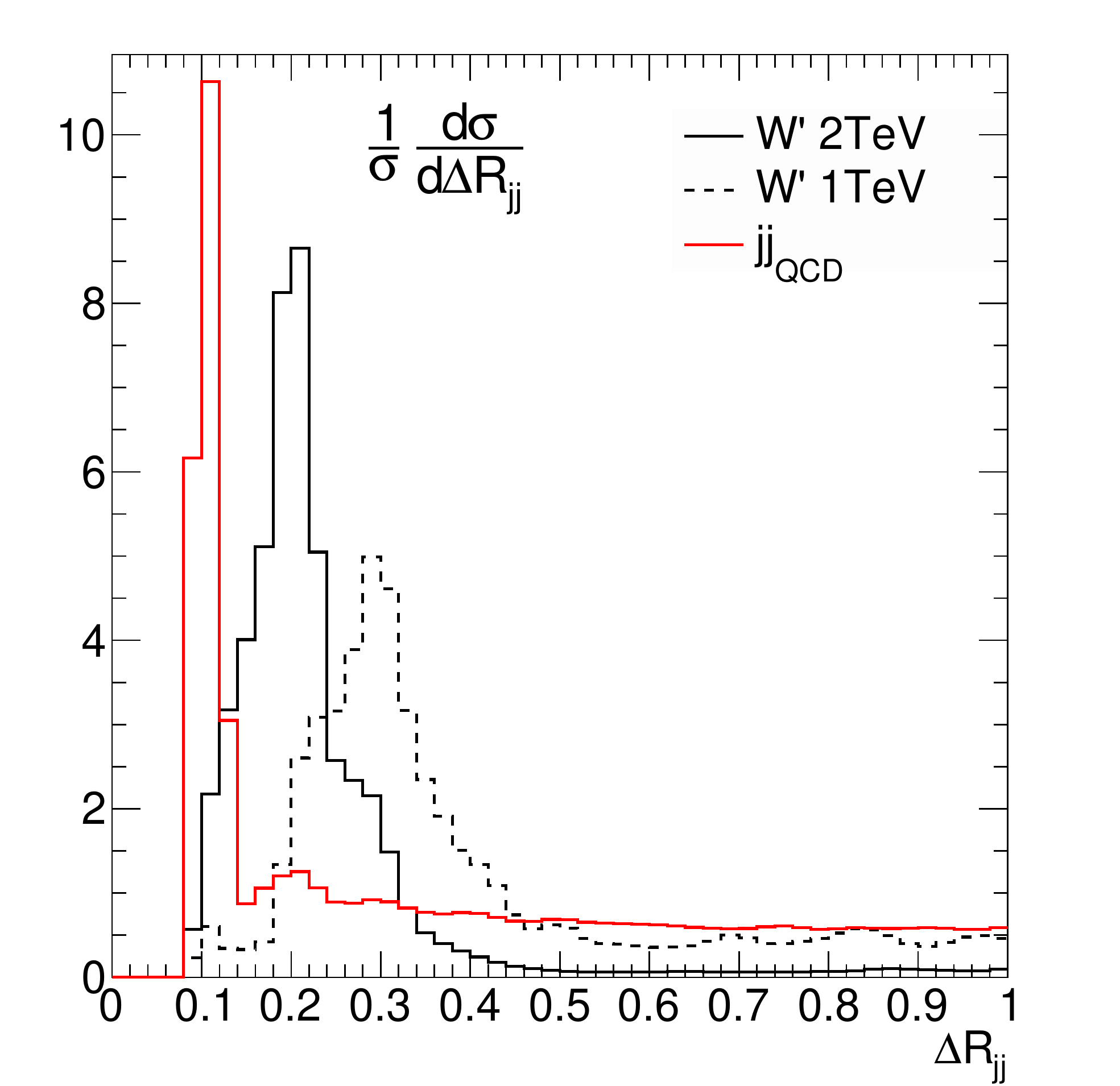}
\includegraphics[width=0.41\columnwidth]{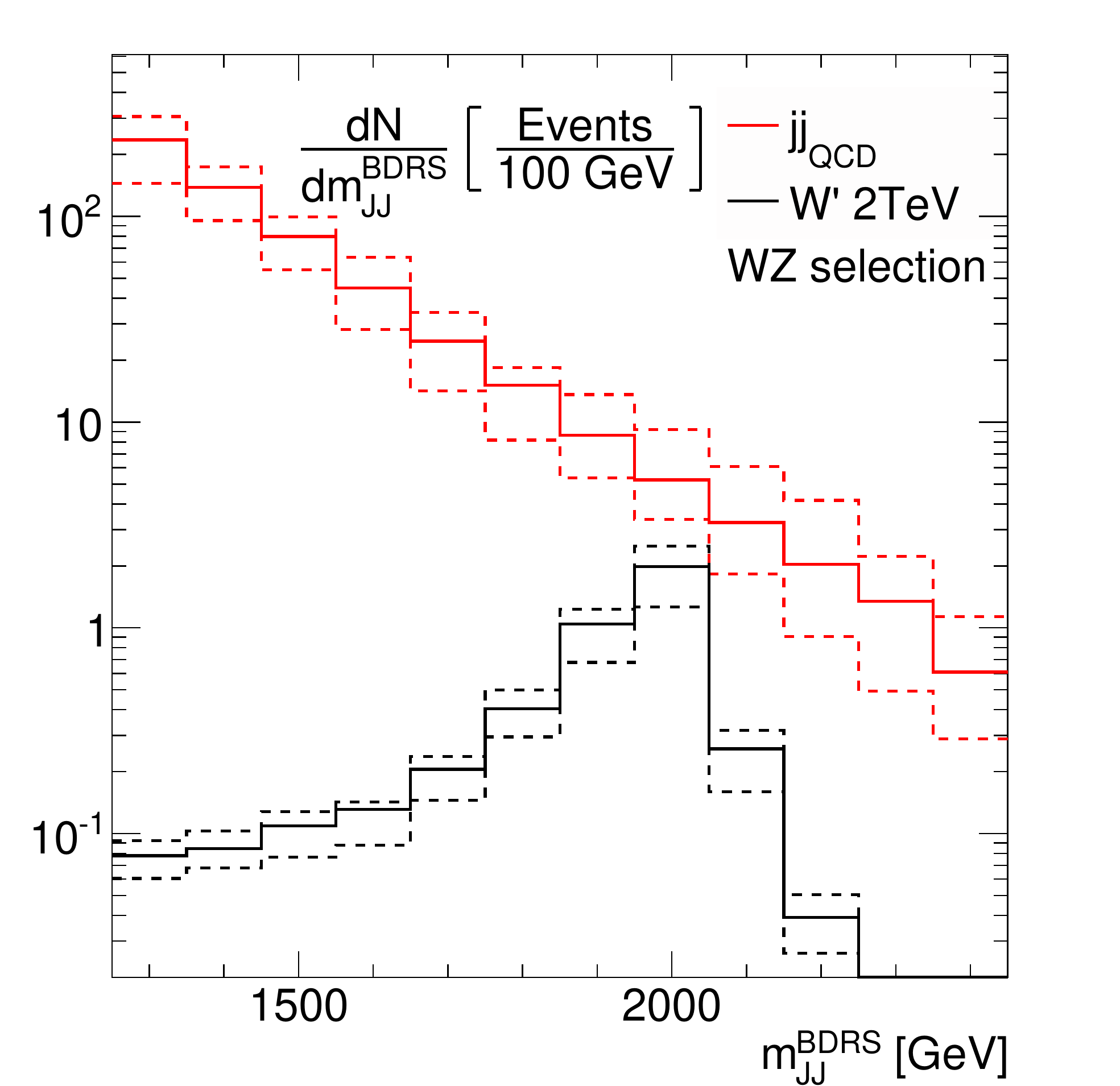}
    \parbox{.8\textwidth}{\caption{Left:  $\Delta R_{jj}$ distribution between the parent jets in the same fat-jet. 
    Right:  Invariant mass distribution $m_{JJ}^{BDRS}$ for the two tagged vector bosons in 
    the WZ selection. The uncertainty bands are obtained by assigning a shift on the $y$ value 
    of $5\%$ upward or downward if the distance of the parent jets in the same fat jet is $R_{jj}>0.5$. If the distance during declustering becomes smaller we parametrise the uncertainty on $y$ by a linear function that  goes from  $5\%$ to $20\%$ for $\Delta R_{jj}$ decreasing from 0.5 to 0.1.
  \label{fig:mjj}}}
\end{center}
\end{figure*}

The way the y-cut is applied in \cite{Aad:2015owa} is different to the scenario envisioned 
for which it was designed. In Fig.~\ref{fig:mjj} (left) we show the $\Delta R$-separation of $j_1$ 
and $j_2$ after the y-cut with $\sqrt{y} \geq 0.2$ is met. We use detector cells of $0.1 \times 0.1$ 
as input to the jet algorithm, for particles or topoclusters based on tracks and the electromagnetic calorimeter as input instead the distribution for QCD di-jet events
would be shifted to even lower values. However, this shows that for the y-cut to be met and the 
declustering to be stopped, for a very large fraction of QCD jets requires to compare the energy ratio
of few or even adjacent topoclusters, i.e the declustering is likely to stop at one of or even the very last merging. As a result, for W/Z and QCD jets the following filtering step with C/A R=0.3 jets is rendered ineffective in reducing the
active area of the jet further. 

On the one hand this seems very desirable, already after meeting the y-cut, pile-up, ISR and UE 
radiation contributing to the jet mass are reduced to a minimum. On the other hand, a source for
large uncertainties is introduced. To our knowledge, the energy-scale uncertainties for topoclusters
are not known. Small jets with large momentum have jet-energy scale (JES) uncertainties of 
$\sim 5\%$ \cite{Aad:2013gja}, but individual topoclusters could have much bigger uncertainties, 
particularly when $\mathcal{O}(1)$~TeV of energy is unevenly distributed over a small number of adjacent topoclusters that are predominantly reconstructed from tracks and the electromagnetic part of the calorimeter. To estimate how an uncertainty of 
the transverse momentum of the subjets, i.e. of the energy of topoclusters, propagates
into an uncertainty on the reconstructed $m_{JJ}$ distribution we shift the $y$-value in Eq.~\ref{eq:ycut} during the declustering procedure and the boson selection by $5\%$ upward or downward if the distance of the parent jets in the same fat jet is $\Delta R_{jj}>0.5$, i.e. an uncertainty corresponding to a calibrated subjet. For smaller $\Delta R_{jj}$ we parametrise the uncertainty on $y$ by a linear function that increases from $5\%$ to $20\%$ while $\Delta R_{jj}$ decreases from 0.5 to 0.1. Hence, we assume that the energy for a large number of topoclusters, corresponding to a $y$-value with large $\Delta R_{jj}$, can be measured more precisely than for few topoclusters. We find that in the invariant-mass region of interest, $1.9~\tev<m_{JJ}<2.1~\tev$, without imposing the gauge boson selection, an upward shift in $y$ of the discussed functional form translates to an upward shift in $m_{JJ}$ of $30\%$ for QCD jets and $20\%$ for $W$ jets. After imposing the gauge boson selection criteria this shift in $m_{JJ}$ increases to $60\%$ for QCD jets while it stays at $20\%$ for $W$ jets, see Fig.~\ref{fig:mjj}  (right panel). As can be expected, Fig.~\ref{fig:mjj} shows that a systematic shift in $y$ that increases for small $\Delta R_{jj}$ mostly impacts on the tail of the $m_{JJ}$ distribution, while leaving $m_{JJ} \lesssim 1800$ GeV much less affected\footnote{If we assume a flat shift for $y$ of $5\%$ ($10\%$), independently of $\Delta R_{jj}$, we find a change in normalisation of $15\%$ ($25\%$) for $1.9~\tev<m_{JJ}<2.1~\tev$ after performing the full $W/Z$ selection.}. For an entirely data-driven background estimate any such sculpturing during reconstruction is dangerous and difficult to consider in the fitting procedure.

\begin{figure*}[hb!]
 \begin{center}
\includegraphics[width=0.41\columnwidth]{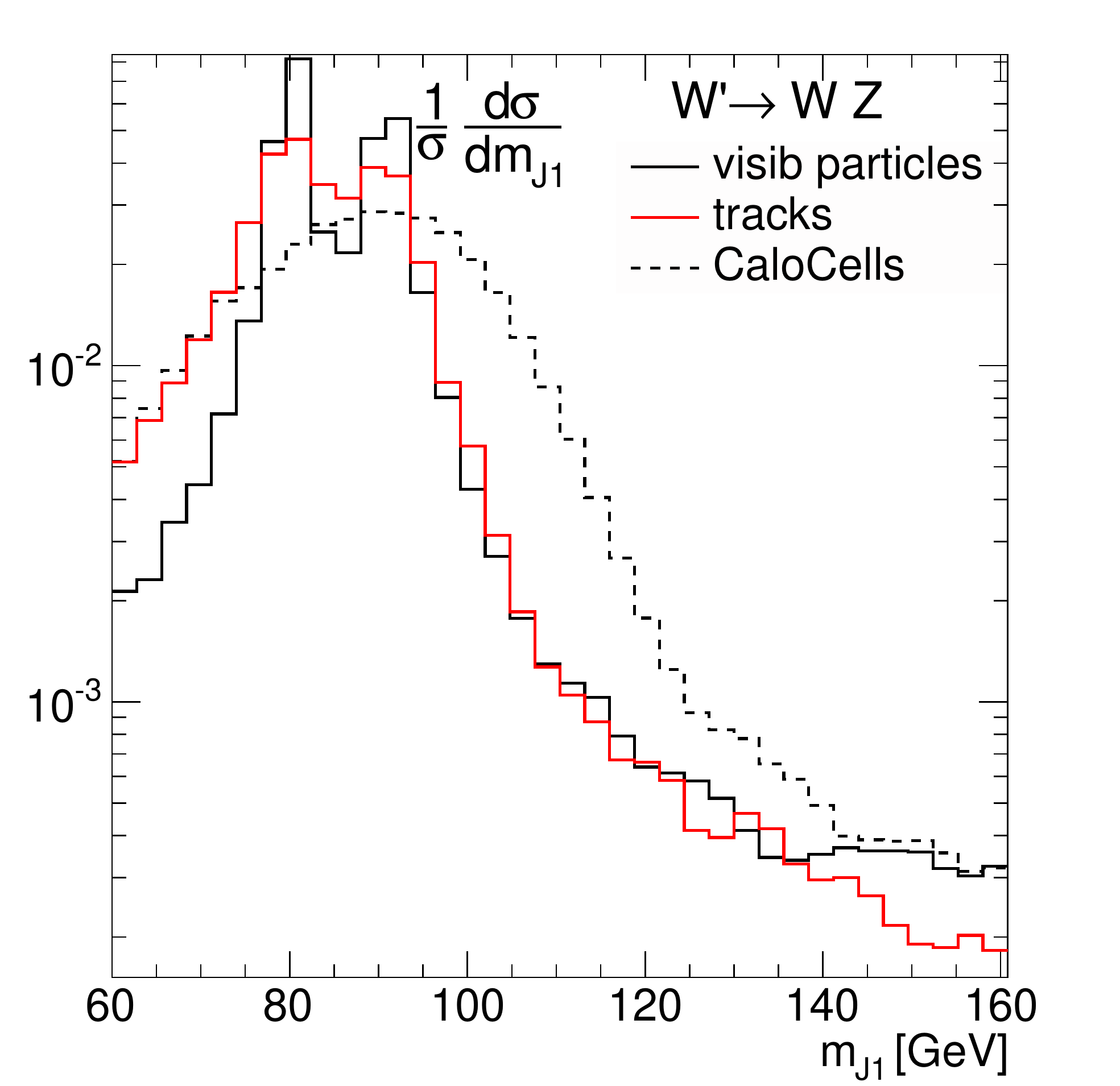}
    \parbox{.8\textwidth}{\caption{Invariant mass distribution for the leading fat-jet $m_{J1}^{BDRS}$ without 
  (visible particles) and with (CaloCells) the calorimeter cell granularity ${\Delta\eta \times \Delta\phi=0.1\times0.1}$. 
  We also display the distribution for the conjunction of calorimeter cells with charged tracks (tracks), where we used the HPTEWBTagger~\cite{Spannowsky:2015eba} for the reconstruction of $m_{J1}$. \label{fig:grooming}
}}
  \end{center}
\end{figure*}

Currently, using calorimeter cells as input for the jets, it is not possible to reliably determine
whether the highly-boosted reconstructed gauge boson is a Z or W boson. The spacial granularity and energy resolution
of the input objects does not allow for a precise mass determination. The black solid curve 
in Fig.~\ref{fig:grooming} shows the reconstructed mass using all visible particles as input, 
while the black dashed line uses massless  ${\Delta\eta \times \Delta\phi=0.1\times0.1}$
calorimeter cells. Adding charged track information in the reconstruction of electroweak 
resonances can be a way forward to improve the sensitivity of the analysis. Dedicated
tagging approaches for W and Z bosons have been designed exploiting the improved 
angular resolution of charged tracks in combination with the energy resolution of the 
calorimeter cells \cite{Spannowsky:2015eba}. The red curve in Fig.~\ref{fig:grooming} 
shows the reconstructed mass distribution after applying the HPTEWBTagger~\cite{Spannowsky:2015eba}
on two C/A $R=0.5$ jets with $p_{T,J} \geq 540$ GeV in each W' event. We find a much 
improved invariant mass distribution, very similar to using all visible particles in the final 
state directly. 

\section{Data Driven Background Estimates}
\label{sec:datadriven}
Data-driven background estimates are becoming increasingly popular within the experimental
community, and, in fact, there are many reasons to believe that they are superior in some
aspects to the usual method relying on theoretical calculations and Monte Carlo simulations.
A textbook example of this method is the discovery of the Higgs boson in the ``golden-plated''
$H\to\gamma\gamma$ channel.  In this channel, there are different sources of backgrounds, 
ranging from the direct production of di-photons in various processes such as 
$q\bar q\to\gamma\gamma$ or $gg\to\gamma\gamma$, which are accessible at different levels 
of theoretical precision, to the misidentification of particles such as $\pi^0$ as photons,
clearly an effect with a size and associated uncertainty driven by the precise knowledge 
of detector performance.  As a consequence of this mix, theoretical methods alone will
not suffice for any meaningful background estimation, and as a consequence, the underlying
di-photon mass spectrum and its uncertainties has been obtained from a fit to data\footnote{
  The impact of its various components, however, has still been carefully checked through
  a combination of highly-precise theoretical calculations and state-of-the-art Monte Carlo
  simulations.  
}.  
The resulting background estimate therefore was taken from this fit and could thus be
subtracted, revealing the ``bump'' of the Higgs boson decaying to two photons at around
125 GeV.  One of the reasons why this worked so brilliantly clearly can be attributed to 
the fact that the fit described the background data in a wide range {\em around} the 
relatively narrow bump with excellent precision.  

In contrast to this example, where data-driven background estimates work 
exceedingly well, many searches focus on the high-energy or high-mass tails of distributions,
effectively the last bins of a distribution.  In such searches, excesses do not manifest
themselves as bumps over otherwise well-understood distributions, but rather as shape
differences in the tails.  As a result, data-driven methods employed there will naturally
rely on an {\em extrapolation of background data} rather than on an {\em interpolation}
as was the case in the discovery of the Higgs boson.  This structural difference clearly
poses challenges to precise background determination in the tails of steeply falling distributions where there is very little lever arm left.  

\begin{figure*}[hb!]
  \begin{center}
    \includegraphics[width=0.47\columnwidth]{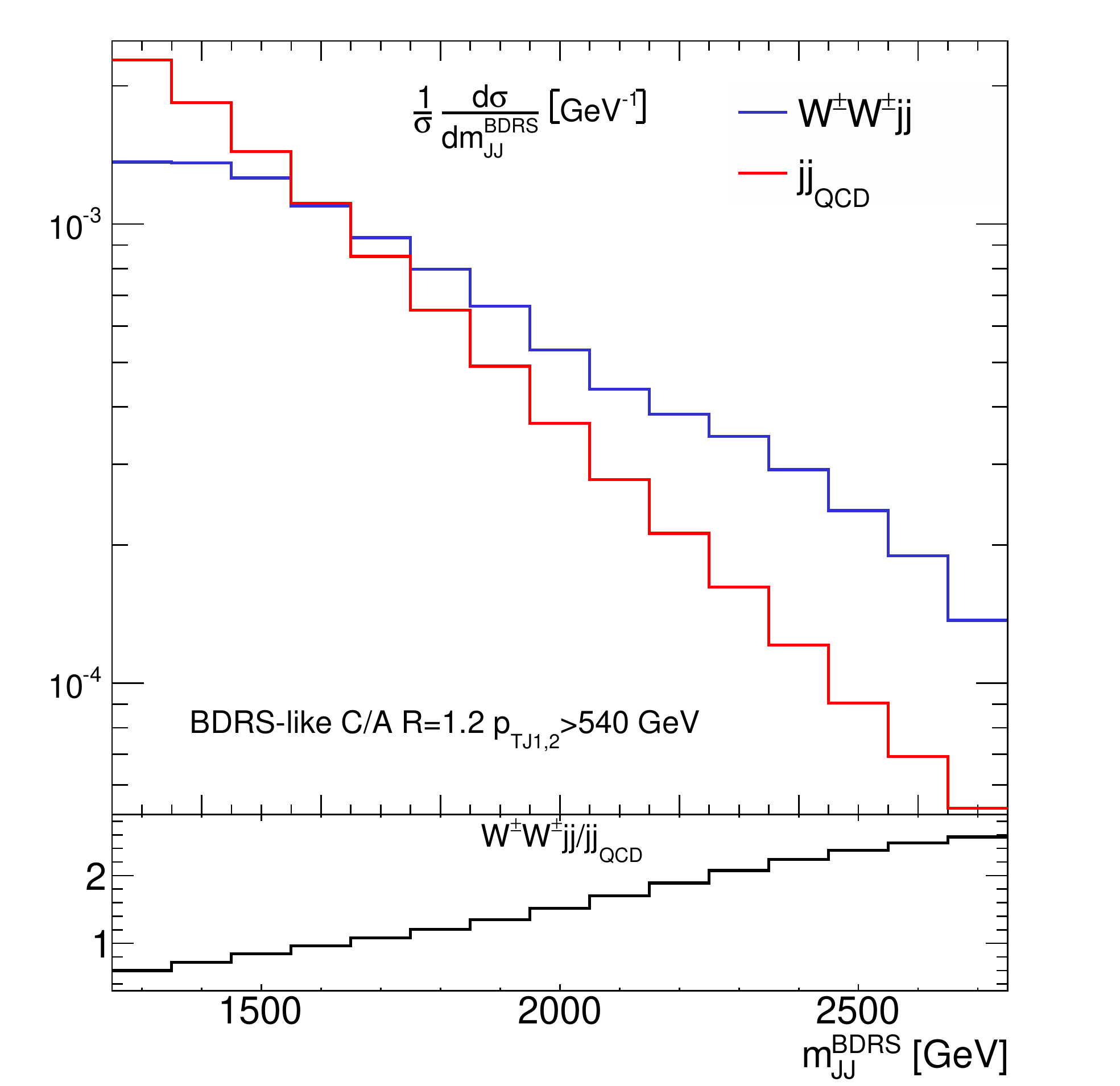}
    \parbox{.8\textwidth}{\caption{
        Invariant mass distribution $m_{JJ}$ for the QCD dijet and same sign W-boson 
        production channels after BDRS grooming and a cut on the transverse momentum of the
        fat jets of ${p_{TJ}>540}$~GeV; note that the distributions here are normalised to the
        relevant cross sections in the kinematic regime in order to exhibit the shape
        difference.  \label{fig:same_sign}}}
  \end{center}
\end{figure*}

There are, broadly speaking, two effects that may introduce subtle issues with the
procedure.  The first, and more obvious one, is that in most modern searches for new
physics the background in fact is a cocktail consisting of different components, {\em i.e.}
different processes.  They can and most often actually will present distinct profiles which 
may induce unexpected structures.  Take, as an example, a case where a very dominant 
background channel -- the one with a much larger cross section in the overall search 
region -- has a profile in some invariant mass distribution that is steeper than the one 
obtained from the subleading channels with the smaller overall cross section.  In the 
case of the ATLAS analysis here, you may think of the production of the di-jet system
through pure QCD as the dominant channel, and the production of pairs of like-sign gauge 
bosons as a relatively suppressed sub-dominant channel, but with a distinctly different
shape in a critical observable, the invariant mass of the di-jet system after BDRS grooming 
and cuts on their transverse momentum.  In Fig.~\ref{fig:same_sign}, we illustrate this 
with the invariant mass distribution for the two channels, where, in order to fully 
appreciate the shape difference, we normalised the distributions on their respective cross 
section in the relevant kinematic regime.  Clearly, in such a case, the data-driven 
background estimate, entirely dominated by the relatively low invariant mass regime, will 
clearly follow the dominant channel, in our example the QCD background.  It then becomes 
a question of actual cross sections of dominant and sub-dominant channels in the relevant 
region, if this shape difference leads to a visible excess with respect to the background 
estimate driven by the dominant channel.  

Such considerations are not elaborated on in the ATLAS publication.  Although some additional 
backgrounds are studied and partially quantified through Monte Carlo simulation, they do not 
appear to enter the data-driven background estimate: Note that 
in the data-driven estimate of ATLAS jet masses in a window around the $W$ and $Z$ masses 
were explicitly omitted, thereby guaranteeing that all EW boson backgrounds did not enter 
the data-driven background estimate at all.  Unfortunately, there is no simulation-based 
discussion of possibly different shapes that may lead to structures in the invariant mass 
spectrum, despite the relatively small total yield of, {\it e.g.}, 6\% for gauge boson pairs. 
It is therefore solely a matter of small cross sections and small statistics that these 
backgrounds do not appear to impact the analysis.  As a consequence the small number of events 
in the high-mass tails renders the validity of the background fit hard to judge.
   
By simulating a large variety of these and other possible backgrounds that did not enter 
any detailed discussion of the analysis, we however confirm that indeed these backgrounds 
do not contribute enough events to explain the observed excess.  In Tab.~\ref{tab:cut_flow} 
we present the cut-flow for all the main SM backgrounds at the leading order. 
Besides the QCD di-jet production, we include 
\begin{itemize}
\item the standard di-boson EW production $ZZ$, $WW$ and $ZW$, named in the paper;
\item top pair production $t\bar{t}$, named in the paper;
\item vector boson plus jet(s) from QCD, $Vj$, named in the paper;
\item the electroweak (EW) production of di-jet systems 
  (for example $q\bar q\to Z^*\to q'\bar{q}'$);
\item $Vjj$ EW production; and 
\item same sign W-boson production $W^{\pm}W^{\pm}(jj)$ in the QCD and EW channel,
\end{itemize}
all simulated using the SHERPA event generator~\cite{sherpa}.
We observe that the extra contributions are suppressed after the complete cut-flow.  
In particular, the relatively large VBF topologies present in some of these backgrounds 
are depleted by the selection $|y_1-y_2|<1.2$, which luckily shape up the extra backgrounds 
more alike the QCD di-jet.  Integrating all the extra components, we obtain $\sim 1$ 
event for $WZ$ selection in the mass range $1.9<m_{JJ}<2.1$~TeV. As these simulations 
were performed only at the leading order, we could easily obtain $\sim 2$ events by 
higher--order effects, which is still far from the cherished excess.

\begin{table}[h!]
  \begin{center}
    \begin{tabular}{|l || c  | c | c  | c |c | c | c| c|}
      \hline
      \multicolumn{1}{|c||}{cuts} &
      \multicolumn{1}{c|}{$W'\rightarrow WZ$} &  
      \multicolumn{1}{c|}{$jj_{QCD}$} &
      \multicolumn{1}{c|}{$t\bar{t}$}&
      \multicolumn{1}{c|}{$VV$}&
      \multicolumn{1}{c|}{$Vj$}  &
      \multicolumn{1}{c|}{$Vjj_{EW}$}  &
      \multicolumn{1}{c|}{$jj_{EW}$} &
      \multicolumn{1}{c|}{$W^{\pm}W^{\pm}jj$} 
      \\
      \hline
      \multicolumn{1}{|c||}{} &
      \multicolumn{8}{c|}{cross sections in fb} \\
      \hline
      \hline
      BDRS $2J$-tag, $p_{T}^J>540$ GeV   & 1.17 & 28302 & 45.6  & 5.34 & 370& 50.8 & 119& 0.50\\
      $\sqrt{y}>0.45$    &  0.59&  4290 &   9.7  & 0.67 &  44 & 5.4 & 10& 0.1   \\ 
      $|y_1-y_2|<1.2$  & 0.45&   2791 &   8.0   & 0.52&   24& 3.2 & 5.8& 0.06\\ 
      $|p_{T1}-p_{T2}|/(p_{T1}+p_{T2})<0.15$  & 0.44 & 2776 & 7.8 & 0.51&  24& 3.2& 5.74& 0.054 \\ 
      \hline
      $WZ$ selection & 0.21 & 26.7 &  0.18  & 0.25 & 0.83& 0.01 & 0.22& 0.0005 \\
      $WZ$ selection, $1.9<m_{JJ}<2.1$~TeV & 0.14 & 0.33 & 0.002  & 0.04& 0.01& 0.0002 & 0.002& 0.00001\\
      \hline
    \end{tabular} 
    \parbox{0.8\textwidth}{\caption{
        Cut-flow analysis for signal and  SM background components. The selections follow
        the ATLAS publication and the cross-sections are given in fb.  \label{tab:cut_flow}
}}
  \end{center}
\end{table}

\begin{figure*}[b!]
  \begin{center}
    \begin{tabular}{cc}
    \includegraphics[width=0.47\columnwidth]{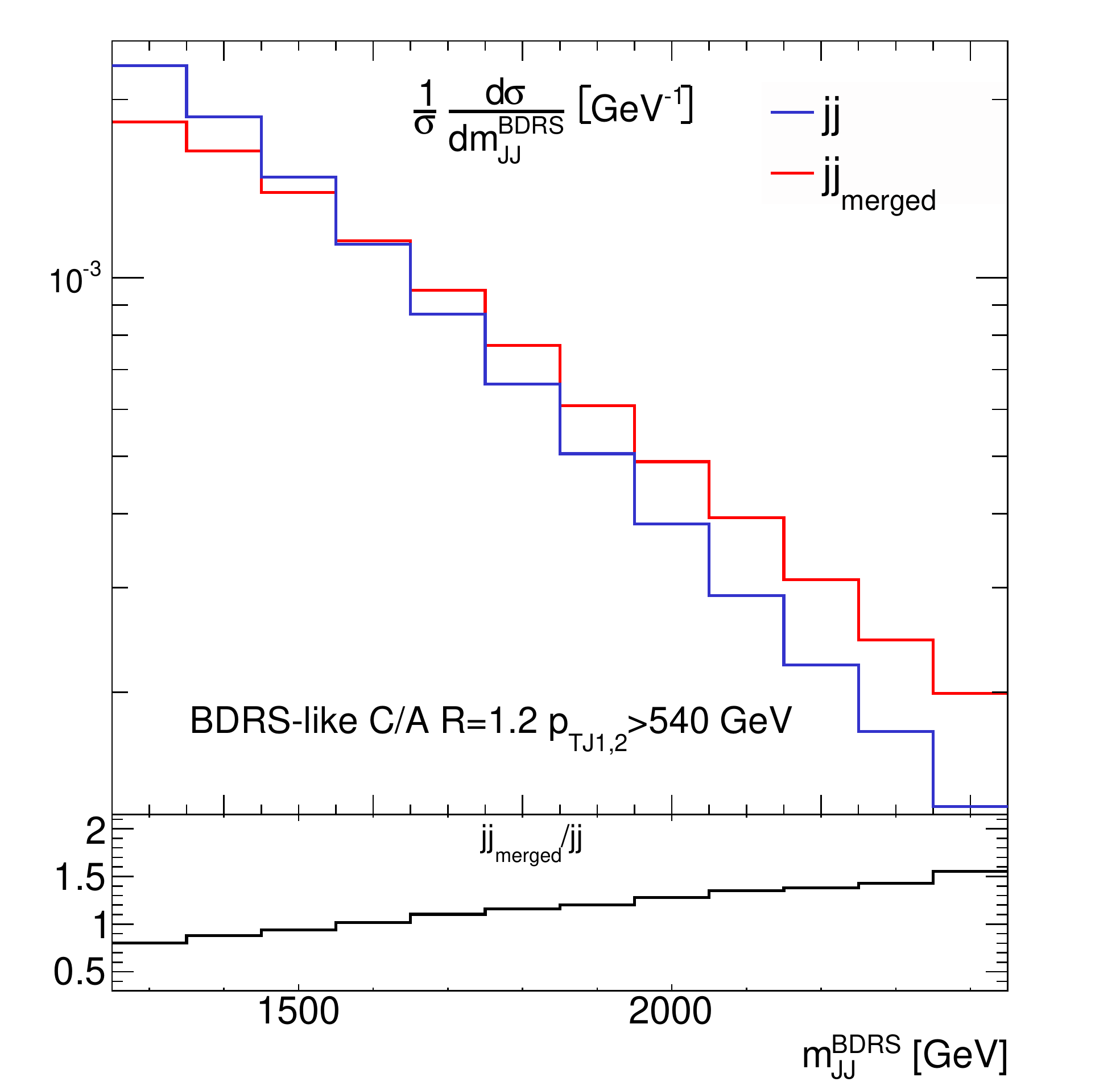} &
    \includegraphics[width=0.47\columnwidth]{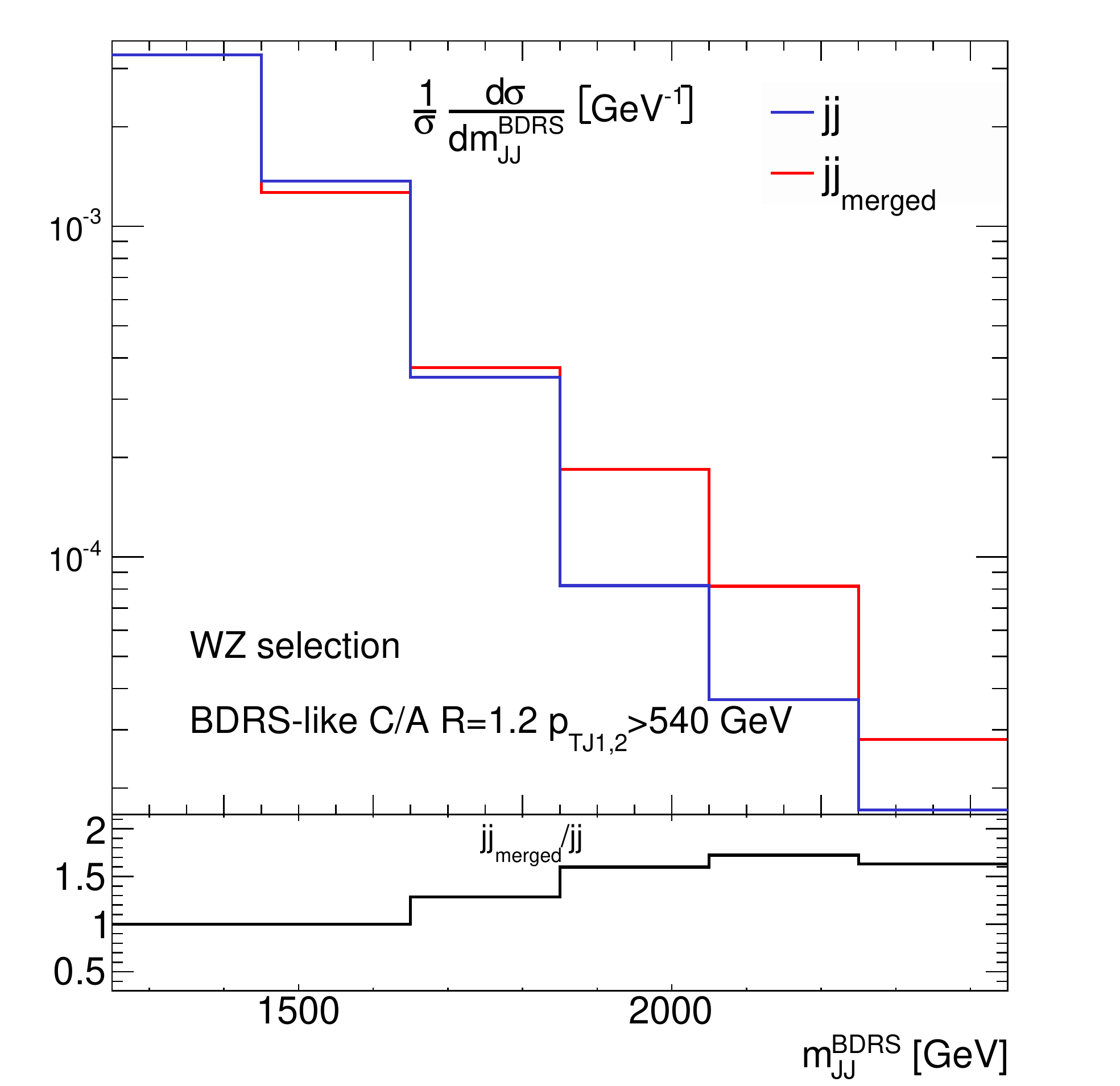} 
    \end{tabular}
    \parbox{.8\textwidth}{\caption{
        Invariant mass distribution $m_{JJ}$ for the QCD dijet production channels after BDRS 
        grooming and a cut on the transverse momentum of the fat jets of $p_{TJ}>540$~GeV.
        Here we compare a simple leading order matrix element sample, augmented by parton
        showering, underlying event and hadronisation with a sample, where multijet merging
        technology has been applied.  In the left panel, the distribution is shown
        before and in the right panel it is shown after additional cuts such as the
        mass condition on the jets.  \label{fig:merging}
        }}
  \end{center}
\end{figure*}
Another possible pitfall is related to the way the dominant sample itself behaves.
Applying jet substructure techniques will always introduce additional scales into 
distributions that otherwise are free of such scales and allow for a smooth fit with
few parameters only.  It is therefore not entirely clear, in how far simple functional
forms of fits are able to capture such multi-scale problems, and as a result this
must be validated, invoking calculations or Monte Carlo simulations.    

In the ATLAS analysis, the data-driven fit was based on the invariant mass spectrum of QCD 
di-jet events before grooming, jet mass requirements etc..  The form of this untagged fit 
was validated by a comparison with Monte Carlo samples from PYTHIA~\cite{pythia8} 
and HERWIG~\cite{herwig}, which have been reweighted to data for untagged jet events.  
The fits have been further validated by looking into the side-bands of jet mass
distributions, in two bins of ${40\,\mathrm{GeV}\le m_J\le 60\,\mathrm{GeV}}$ and
$110\,\mathrm{GeV}\le m_J\le 140\,\mathrm{GeV}$.  From the paper it remains unclear 
whether this treatment results in different fits, which are then interpolated into
the signal region of $60\,\mathrm{GeV}\le m_J\le 110\,\mathrm{GeV}$, or if there 
is only one fit that captures all masses.  
  
The Monte Carlo samples used for the fit validation are based on leading order matrix 
elements, producing the two jets, supplemented with the parton shower.  Looking for 
substructures in fat QCD jets on the other hand is sensitive to those topologies where at 
least one splitting was hard enough to give rise to two distinct subjets, which are 
possibly better described by multijet merging methods~\cite{merging}, 
and which have routinely been used in LHC analyses at RUN I.  Applying such methods, again
using SHERPA, indeed appears to give a result that differs from the parton-shower only
approach to jet substructure, see Fig.~\ref{fig:merging}.  There we have compared
both approaches for jets after BDRS grooming and with a jet $p_T$ cut of $540$ GeV, but 
before applying all other cuts of the analysis.  These cuts, and in particular the 
cut on the jet masses to be around the gauge boson masses, 
$60\,\mathrm{GeV}\le m_J\le 110\,\mathrm{GeV}$, appear to even enhance the difference. 
While this may not change the functional form of the fit, it is obvious that it would change
the parameters in a way that will most likely be very sensitive to the mass cut applied.

\section{Outlook}
\label{sec:conc}

In this publication we tried to better understand the analysis presented by ATLAS 
in~\cite{Aad:2015owa}, where technologies for highly boosted objects have been
combined with data-driven background estimates in the search of heavy resonances
decaying into pairs of weak gauge bosons.  By scrutinising the analysis from a more
theoretical perspective we found a number of issues that may warrant a more detailed
discussion, especially in light of the upcoming searches for new physics at Run II
of the LHC:
\begin{itemize}
\item First of all, we would like to stress that it is important to adjust the
  parameters of substructure analyses on fat jets to the process and kinematic regime 
  being considered.  For example, for a Higgs boson with a mass of $m_H=120$ GeV and a
  transverse momentum of $p_T\ge 200$ GeV, a typical fat jet size would be given
  by $R\approx 2m_H/p_T \approx 1.20$, a value used in the original BDRS paper.
  For electroweak gauge bosons with a mass of $m_W=80$ GeV and a transverse
  momentum of $p_T\ge 540$ GeV, however, a suitable fat jet radius would 
  probably more of the order of $R\approx 2m_V/p_T\approx 0.4$ -- picking a
  larger radius thus merely increases the probability to pick up subjet structures
  that did emerge from additional QCD particle production, for example from
  initial state radiation, the underlying event or even pile-up instead of from
  the decay of the gauge boson.
\item In addition, it is clear that a non-negligible fraction of subjets will
  overlap and sometimes even hit the same area of the detector, which naively
  can be assumed to have a granularity of $0.1\times 0.1$ in the $\eta$-$\phi$
  plane.  As a consequence the additional filtering with Cambridge--Aachen jets 
  of the size $R=0.3$ does probably not add any discriminating power to the 
  analysis, but rather obfuscates it.  This is because it is highly likely
  that by demanding two such jets one of them will contain both decay products
  of the gauge boson and the other jet will therefore originate from some
  generic QCD noise.  As a result, the invariant mass of these two jets will
  not be able to serve as a meaningful signal for the gauge bosons.
\item There is yet another issue related to using sub-jet structures, namely
  the question in how far uncertainties in their momentum/energy scale are
  fully understood.  Unfortunately the paper gives systematic uncertainties for 
  the overall jet $p_T$ scale and resolution and the jet mass scale for
  the background only, while discussing other, crucial effects related to
  grooming and filtering only for the signal.  When trying to naively estimate
  these effects also for the background we found relatively large 
  uncertainties, which seem to be partially due to the granularity of the 
  calorimeters.  Adding these uncertainties to the analysis reduced the significance
  of the excess dramatically and can even shape the tail of the $m_{JJ}$ distribution. 
  The observed effect is also concerning in light of future applications of jet substructure
  tools, in particular when methods are used that rely on the direct use of objects with
  potentially large energy scale uncertainties, i.e. individual topoclusters or particle-flow objects.
\item In order to overcome these problems and to allow for a jet filtering
  with a finer granularity we suggest to also rely heavily on tracking information.
  Some naive, preliminary analysis seems to indicate that this is an avenue
  which is worth further studies.  For the reconstruction of even heavier resonances
  using tracks will become unavoidable.
\item In addition we want to challenge the way this analysis and possibly others 
  rely on data-driven background estimates.  While after some additional checks many
  of our initial concerns have proven to be inconsequential, there are still a 
  few issues we would like to raise.  One of them is related to the fact that
  some backgrounds, such as like-sign $W$ production or VBF-type topologies for
  gauge boson pair production apparently have not been considered, while 
  others, like ``ordinary'' QCD driven gauge boson production have been discarded
  based on a relatively low overall yield.  This of course implicitly assumes
  that such sub-leading backgrounds behave in a sufficiently similar way with
  respect to the leading ones such that they do not introduce shapes in relevant
  distributions.  This check is missing in the publication.  

  In fact, the data-driven background estimate appears to be avoiding exactly
  such processes, due to the choice of side-bands in the jet masses and it is thus
  unclear from the paper in how far these backgrounds contribute.  We therefore
  chose to check their effect and thereby explicitly confirm a finding that was at
  best implicit in the publication.  
\item Finally, we would like to draw attention to the fact that for QCD backgrounds
  in sub-jet analyses, the parton shower alone may not be the optimal tool.  We
  suggest that in future analyses multijet merging techniques are being used, as
  they are better able to capture splittings inside a fat QCD jet that are hard
  enough to produce two energetic subjets with a sizable relative transverse
  momentum.
\end{itemize}

\section*{Acknowledgements}
We would like to thank various members of the ATLAS collaboration, in particular Alex Martyniuk, David Miller and Enrique Kajomovitz,
for fruitful discussions that helped us to better understand some of the intricacies 
of their analysis.
We are grateful to  the  UK  Science  and  Technology  Facilities  Council STFC for
partially funding our research, F.K.\ and M.S.\ also acknowledge financial support 
by HiggsTools ITN under grant agreement PITN-GA-2012-316704.  
F.K.\ acknowledges additional support by the ERC Advanced Grant MC@NNLO (340983).
This research was supported by the Munich Institute for Astro- and Particle Physics (MIAPP) of the DFG cluster of excellence "Origin and Structure of the Universe".


\end{document}